# What is System Dynamics Modeling? Defining Characteristics and the Opportunities they Create

Asmeret Naugle[1], Saeed Langarudi[2], and Timothy Clancy[3]

[1] Sandia National Laboratories, abier@sandia.gov
[2] University of Bergen
[3] Dialectic Simulations Consulting, LLC

**Abstract**

A clear definition of system dynamics modeling can provide shared understanding and clarify the impact of the field. We introduce a set of characteristics that define quantitative system dynamics, selected to capture core philosophy, describe theoretical and practical principles, and apply to historical work but be flexible enough to remain relevant as the field progresses. The defining characteristics are: (1) models are based on causal feedback structure, (2) accumulations and delays are foundational, (3) models are equation-based, (4) concept of time is continuous, and (5) analysis focuses on feedback dynamics. We discuss the implications of these principles and use them to identify research opportunities in which the system dynamics field can advance. These research opportunities include causality, disaggregation, data science and artificial intelligence, and contributing to scientific advancement. Progress in these areas has the potential to improve both the science and practice of system dynamics.

# Introduction

The field of system dynamics needs clear definition. We begin that process here by characterizing the quantitative modeling portion of the field; systems thinking and other qualitative methods are not considered here. A shared understanding of the field's defining characteristics can clarify what system dynamics modeling is and what it is not, helping to strengthen the impact of the field and facilitate its progress.

Without a shared understanding of the defining characteristics of system dynamics, the field's identity is often characterized by common methods or historical capabilities, both of which restrict potential for growth by focusing on the past. When the field is defined too narrowly, we restrict innovation; when the field is defined too broadly we struggle to understand its impact, educate newcomers, and recognize appropriate uses of system dynamics. The field is also in need of a consistent identity; Homer (2013) discussed an identity crisis in the system dynamics community, and Sterman (2018) noted that contradictory definitions have led to a state of ambiguity and "a confusion about what system dynamics is" (Sterman, 2018). System dynamics was originally developed in a world without much computational power, with relatively few modeling methods available, and without modern capabilities like artificial intelligence that might boost the field's potential. Technological progress has changed the landscape of analytic options, further strengthening the need for a common identity. We believe that a clear identity based on defining characteristics can improve system dynamics' practical advantage and scientific impact while maintaining the core philosophy that binds the field together.

We propose that system dynamics modeling is defined by the following characteristics:

1. Models are based on causal feedback structure
2. Accumulations and delays are foundational
3. Models are equation-based
4. Concept of time is continuous
5. Analysis focuses on feedback dynamics

All models with these characteristics should be considered system dynamics. Other techniques involved in the system dynamics process, such as systems thinking and group model building, are important to the field but not considered here; in the remainder of the text when we use word modeling we are referring only to models and processes that include quantitative aspects.

In following sections we provide historical context for the defining characteristics, discuss the characteristics themselves, and identify opportunities for growth of the field, calling for new development that holds true to the core defining characteristics of system dynamics modeling. We invite and encourage a larger discussion on these defining characteristics and the opportunities for growth that they present.

## Previous attempts to define the field of system dynamics

The field of system dynamics has struggled to define itself in a consistent way. Early work by Jay Forrester, the field's founder, defined system dynamics as "the science of feedback behavior in social systems" (Forrester 1971, p. 400), "a philosophy of structure in systems" (p. 406), and "a body of principles that relate structure to behavior" (p. 400). Attempts at defining system dynamics have generally fallen into three categories: definitions that focus on the structural components of system dynamics models; system dynamics as a methodology; and definitions that place system dynamics within broader paradigmatic contexts.

One early definition based on structural components described feedback loop as the "basic building block" of system dynamics, with the substructure made up of levels (later known as stocks) and rates (later known as flows) (Forrester 1971, p. 72-74). Two more components were eventually added to the structural toolbox, resulting in stocks, flows, auxiliaries, and constants being considered the 'constitutive structures' of system dynamics models (Pruyt 2013, p. 85–87), with underlying equations further defining the models (p. 35). Structural definitions are sometimes used to compare and contrast system dynamics with other simulation paradigms, such as agent-based modeling and discrete event simulation (Dooley, 2017, p. 829). Richardson (2011) highlights how all of these structures exist within closed boundaries representing the system of interest, and argues that this endogenous point of view is the foundation of system dynamics. Yasarcan (2023) focused on dynamic complexity specifically, identifying accumulations, feedback loops, nonlinearities, and delays as its four main elements.

Within the field, system dynamics is often defined as a methodology. For instance, Ford (2010, p. 7) defines system dynamics as "a methodology for studying and managing complex systems that change over time", while Sterman (2000, p. 4) asserts that "system dynamics is a method to enhance learning in complex systems." Saeed (2017) defines system dynamics as a set of methods for developing a reference mode, a dynamic hypothesis, a structurally and behaviorally valid model, and policy design, with a binding focus on realistic policy analysis. As technical progress has advanced, some widely encompassing descriptions of the methodological boundaries of system dynamics have emerged. For example, Sterman (2019) asserted that "system dynamics models can be and are instantiated using a variety of modeling architectures including compartment models, individual (agent-based) models, dynamic network models, discrete simulation models and so on."

Other definitions have focused on positioning system dynamics in broader paradigmatic contexts. Lane, for example, discusses system dynamics in the context of social systems theory, suggesting that system dynamics is grounded in the integration of human agency and system structure theories (Lane, 2001a; Lane, 2001b). Pruyt (2006) discusses system dynamics within the context of scientific philosophy, defining the field specifically as a paradigm (in contrast to a philosophy or methodology) and noting its close relationship with critical pluralism.

While these historical definitions are all useful, none of them have satisfied the field's need for a coherent identity. We need a definition that explains which models are considered system dynamics and why, while allowing room for change and innovation. Structural definitions are useful for identifying traditional system dynamics models, but they do not capture the core philosophy of the field and do not allow for new developments. Methodological definitions have a similar problem; they focus on practices that have proven historically useful, but methods are moving targets as capabilities advance over time (Sterman, 2018). Paradigmatic context definitions are useful for explaining how system dynamics interacts with other philosophies and fields, but do not provide clarity or a coherent identity for system dynamics itself.

The definition discussed here takes a different approach, capturing core characteristics that are both essential to the conceptualization and practice of system dynamics and also distinguish system dynamics from other modeling paradigms. We are not proposing broader paradigmatic conclusions or delving into epistemological or other philosophical issues, and are not delving into qualitative modeling. Instead, we find value in a simple, clearly articulated, short list of the defining characteristics of system dynamics models. Epstein (2005) articulates this type of definition for agent-based modeling, focusing on the characteristics of heterogeneity, autonomy, explicit space, local interactions, bounded rationality, and non-equilibrium dynamics. We find such a definition to be useful in capturing the essence of the field while allowing for evolution and innovation, and we present the system dynamics modeling equivalent here.

## The defining characteristics of system dynamics modeling

These defining characteristics were selected through a deductive process informed by the literature to (1) capture the core philosophy of system dynamics, (2) describe the theoretical and practical principles that make system dynamics modeling unique, and (3) apply to historical work but be flexible enough to hold as technical capabilities progress.

### *Characteristic 1: Models are based on causal structure with feedback*

System dynamics examines the world through the lens of causal feedback loops. Model structures are based on relationships between elements in the system. These relationships are specifically causal; other relationships, such as those that are simply correlational, are generally not included. We seek a holistic approach, attempting to consider all the relevant causal relationships defined by the focus problem, with the problem defined by a model boundary (Richardson, 2011), granularity, and pertinent variables. The causal relationships within this system are arranged into interconnected causal structures, often displayed as causal loop diagrams (see Bala et al. 2017 for more information on developing causal loop diagrams).

Causal structures used in system dynamics always involve feedback loops. Feedback loops, which are known in some other fields as cycles or circuits, occur when these causal relationships chain together in such a way that a full loop is formed. In other words, the causal structure of the system is connected in such a way that a change to a focus variable will cause a chain of events that eventually comes back to influence the focus variable itself. Feedback loops are either positive (reinforcing) or negative (balancing). Positive feedback loops push a system in one accelerating direction, leading to exponential growth. Negative feedback loops balance these systems, tending to pull behavior toward equilibrium. The causal structure of a model, including the specific combination of feedback loops, determines how its output behavior unfolds over time.

*Characteristic 2: Accumulations and delays are foundational*

System dynamics considers accumulation, and the associated concepts of delays and inertia, to be essential in determining real-world dynamics (Forrester, 1971). All system dynamics models include stocks, or variables in which something accumulates, and flow variables that determine changes to those stocks. Stock and flow structures introduce a realistic path dependence to system dynamics models. This accumulation worldview separates system dynamics from modeling approaches with differentiation worldviews. According to Forrester (1968, p. 411), the differentiation worldview (the use of derivatives or differential equations to explain a system's dynamics) "tends to focus attention on the wrong direction of causality" by emphasizing rates of change and minimizing the role of stocks, while the accumulation worldview highlights the role of the stock and how it drives dynamic behavior.

The accumulation worldview is fundamental for providing consistency and coherence between different layers of philosophy, theory, and practice of system dynamics. Any feedback loop in a system dynamics model must include at least one stock; without the inertia provided by a stock the existence of a feedback loop would indicate that the stock has multiple simultaneous values, which is impossible.

Delays are a key consideration of system dynamics that follows directly from our attention to accumulation. Delays are caused by accumulation processes with asynchronous inflows and outflows. Delays play an essential role in the material and information processing involved in behavioral, social, and natural phenomena.

*Characteristic 3: Models are equation-based*

Each variable in a system dynamics model is defined by a mathematical equation. This equation describes how the variable will change as its causal influences change. This contrasts with other modeling paradigms in that nearly all elements of a system dynamics model (excluding model support elements such as those for model testing or policy evaluation) are intended to describe realistic mechanisms. Our models are simulated by calculating the value of each variable at a starting time step, and then updating the values for all variables at the next time step, and so on.

In this way, system dynamics models simulate the logical consequences of known assumptions, with these assumptions made concrete and comparable as model equations. This focus on concrete equations makes system dynamics models relatively reproducible, comparable, and easy to use. It also facilitates policy analysis, since the causal mechanisms explicit in model equations allow us to study leverage points in a system that might lead to outcomes of interest.

Model equations are used to represent and connect the four building blocks of a system dynamics model: stocks, flows, auxiliaries, and constants. As discussed above, system dynamics focuses on accumulations, which are represented in our models as stock and flow equations. This architecture can easily translate to integral equations, connecting calculus philosophy to system dynamics practice. Auxiliaries represent variables that change but do not involve accumulation. Constants do not change but can be used in the equations that define the other building blocks. Stochastic variables and other statistical techniques, as well as discrete events, can be incorporated into a system dynamics model, but they tend to be and used sparingly and for specific purposes.

*Characteristic 4: Concept of time is continuous*

System dynamics uses a relatively realistic concept of time. We assume that time is continuous and pay close attention to how the behavior of a system evolves over time. This consideration of time allows system dynamics to model accumulations, which facilitates the examination of feedback loops and their effects on a system's behavior. It also aligns well with the real world; time marches on, with events and accumulations occurring at their own intervals. Decisions may be made with the best information available at a given time but implemented in the context of future conditions. Time delays affect accumulations, which might have major consequences to a system's dynamics. Problems and their associated system dynamics models generally focus on some bounded time scale, although dynamics at different time scales can interact.

Practically, system dynamics models are simulated by segmenting the conceptually continuous time horizon into discrete time steps. While a system dynamics model represents a continuous system defined by integration, the associated simulation represents the system discretely by approximating the model with difference equations. The consideration of continuous time contrasts with modeling paradigms with more discrete philosophies; for example, some other fields simulate the forward movement of time based on events or ticks, rather than using a continuous time horizon to provide context for a system's dynamics.

*Characteristic 5: Analysis focuses on feedback dynamics*

System dynamics is fundamentally concerned with studying how the causal feedback structure of a system drives that system's behavior. As discussed above, the field looks at dynamic behaviors through a focus on causal feedback loops that incorporate accumulations and delays expressed as mathematical equations based in continuous time. In this way, structure is linked to behavior, and potential leverage points that might adjust behavior are identified. System dynamics generally uses simulation to facilitate this analysis, although non-simulation techniques do exist; for example, analytic techniques have been used to investigate structural dominance of system dynamics models (Kampmann and Oliva, 2008). Various methods exist for conducting feedback analysis to connect structure to behavior.

In contrast, other simulation paradigms focus on different goals. For instance, agent-based modelers focus on emergent behavior resulting from agent attributes, agent behavior, and agent-agent or agent-environment interactions (Railsback and Grimm, 2019); network analysis focuses on specific combinations of edges and nodes in a network (Newman, 2018); discrete event simulation focuses on sequences and processes (Fishman, 2001). System dynamics' focus on feedback analysis is one of the things that distinguishes us from other fields, thus providing a clear identity and character rooted in our tradition of rigorous scientific inquiry.

## Opportunities for growth in the system dynamics field

By comparing system dynamics modeling's defining characteristics with the field's traditional focus areas, we can investigate gaps. These gaps highlight opportunities for the system dynamics research community – new directions that hold promise for work that remains philosophically system dynamics but extends the field's potential for impact. This discussion highlights a few of these opportunities. Each discussion includes a list of initial research questions that the system dynamics field should address. This list is by no means exhaustive, and we encourage further exploration of potential new frontiers for system dynamics.

*Causality*

Causal relationships define the feedback structures that all system dynamics is based on (defining characteristic 1), but the field has not put sufficient effort into investigating causality itself. The field needs more focus on what causality means in the context of system dynamics, how to clarify the validity of our causal assumptions, and how system dynamics might contribute to causal discovery and causal inference more broadly. Sterman (2018) advocates using statistical causality methods in system dynamics; we take this a step further, suggesting a research thrust within the system dynamics field to determine the best ways to address causality in system dynamics modeling.

Since causal structure is so fundamental to system dynamics, confidence in a model requires confidence in its causal assumptions. Barlas (1996) discusses methods for validating a model's structure but focuses on face validation of equations with experts, rather than evaluating the causal relationships themselves. Recent advances in the statistical causality literature might provide opportunity to reinvigorate this discussion, with new ways of approaching causal discovery and causal inference (Nogueira et al., 2022). Some of these methods rely on directed acyclic graphs (DAGs), but others allow room for "cycles", or feedback loops. The system dynamics field should investigate the utility of these methods to our models, and how the methods might be adapted based on the field's defining characteristics. We also need approaches for dealing with the fact that the causal structures driving real-world system behaviors may not be static. To fully understand the validity of a model's causal structure, we need to investigate the circumstances under which that structure maintains that validity. The data science field uses the notion of *concept drift* (Lu et al. 2018) to address this issue; the system dynamics field needs a similar line of research.

Not all models require full evaluation of the validity of causal structures. In some cases, a model is designed to present a yet-to-be-tested theory or a particular expert's point of view. These models can be useful but should be presented carefully. For example, a model intended for storytelling may require limited investigation of causal validity, *as long as* the model's purpose is identified as such and its uses are appropriately constrained. Nevertheless, system dynamics should not be used to make unsupported points; observers can be misled when models are presented without discussion of the source of, and confidence in, the causal structure. The system dynamics field needs a clear categorization of intended purposes of models, and the levels of scrutiny those purposes require.

Research questions on causality in system dynamics include:
- *How can we validate, or build confidence in, the causal structures used in system dynamics models?*
- *How do we determine the circumstances (time horizons, situational changes, etc.) under which a causal structure holds?*
- *Can system dynamics contribute to broader discovery of causal relationships?*
- *What characteristics of system dynamics projects determine whether investigation of causal validity is needed?*

*Disaggregation*

System dynamics can be used to study systems at various levels of aggregation. Historically the field has focused on relatively aggregated models. While some problems are well-suited for analysis with small models (Ghaffarzadegan et al. 2011), this focus on aggregation was at least partly due to computational limitations that no longer apply. Aggregated models also align well with our focus on accumulations (defining characteristic 2). But accumulations are useful for modeling a wide variety of phenomena, including things that drive behavior in disaggregated systems, such as decision making. System dynamics is sometimes contrasted with agent-based modeling; since agent-based modeling requires disaggregation, system dynamics is sometimes wrongfully contrasted as the aggregated modeling paradigm. Nevertheless, aggregation does not define system dynamics, which leaves the field open to pursue opportunities at various levels of aggregation and to learn how the principles of system dynamics apply to disaggregated systems. Sterman (2018) makes a similar call for considering disaggregation in system dynamics; however, he focuses on thinking of traditional agent-based models as a subset of the system dynamics field, whereas we suggest using system dynamics and its defining characteristics to investigate disaggregated systems.

Exploring the use of disaggregation in system dynamics modeling might broaden the types of problems we can help to solve. Different problems are best studied at different levels of aggregation, with some problems requiring analysis at multiple levels of aggregation. For example, understanding the influence of behavior on COVID-19 dynamics requires asking questions from both the aggregated and disaggregated perspectives. Rahmandad et al. (2021) used an aggregated system dynamics model to look at the influence of behavioral factors such as adherence fatigue on the spread of COVID-19, while Naugle et al. (2022a) use a disaggregated system dynamics model to evaluate how conflicting information about COVID-19 traveled through social networks to alter spread. Some work has been done looking at the interconnection of system dynamics and agent-based modeling, mostly focused on comparing the paradigms (for example, Borshchev et al., 2004; Rahmandad and Sterman, 2008), with a few advancing that discussion toward hybrid modeling (for example, Swinerd and McNaught, 2012; Martinez-Moyano and Macal, 2016; Anderson et al., 2018; Langarudi et al., 2021). However, most hybrid modeling has focused on connecting agent-based models to aggregated system dynamics models; disaggregated system dynamics models themselves remain understudied.

Research questions on disaggregation in system dynamics include:
- *What applications might benefit from disaggregated system dynamics analysis?*
- *How should problem characteristics drive the selection of aggregation level?*
- *What model analysis methods, existing or new, should be used with disaggregated models?*
- *Do common principles and practices of system dynamics apply to disaggregated systems, or do they need to be adapted?*

*Data science and artificial intelligence*

Since the introduction of system dynamics, the field of data science has flourished. Some new data science capabilities might improve system dynamics' causal discovery (defining characteristic 1), equation development (defining characteristic 3), and analytic (defining characteristic 5) capabilities, if not more. The system dynamics field has begun to take advantage of advances in data science, but significant opportunity exists (Pruyt et al., 2014).

Making the best use of available data, including big data, could boost the impact and utility of system dynamics overall, and has become expected in the current research environment. The system dynamics field sometimes views itself as a counterpoint to big data options; system dynamics does not require abundant data to be useful, and can rely on expert opinion in situations where data does not exist. The integration of system dynamics and data science has huge potential benefits, both for improving the rigor of system dynamics and for improving interpretability of data science.

Classical statistical methods are certainly still useful for model development, but machine learning methods can facilitate more generalized discovery of predictive patterns from data (Bzdok, 2018). Machine learning techniques have recently been incorporated into system dynamics model parameterization, calibration (Chen et al., 2011), and analysis (Ozik et al., 2016; Pruyt and Islam, 2016; Edali, 2022), although significant room for innovation exists. The system dynamics field should continue to incorporate modern data science techniques into all aspects of the system dynamics modeling process.

System dynamics modeling may also contribute to novel artificial intelligence (AI) capabilities. For example, simulations can be used as testbeds for training AI and studying its utility (Naugle et al., 2022b; Lakkaraju et al., 2022). Some early efforts have been made to use AI for automated causal model generation (Schoenberg, 2019; Rackauckas, 2020). Domain-inspired machine learning techniques (Baker et al., 2019), which aim to incorporate theory and other knowledge into machine learning algorithms, might also benefit from system dynamics modeling. These recent examples can highlight opportunities, but many potential directions have not yet been attempted; this area of opportunity deserves significant brainstorming and exploration, and new possibilities will emerge as AI methods evolve.

Research questions on data science and artificial intelligence in system dynamics include:
- *How should system dynamics incorporate data science methods into model parameterization, calibration, and analysis?*
- *Are there novel ways the data science methods can improve system dynamics?*
- *Can system dynamics contribute to novel AI methods?*

***Contributing to scientific advancement***

System dynamics attempts to link a system's causal feedback structure (defining characteristic 1) to its behavior (defining characteristic 5), which may facilitate new contributions to scientific understanding. The system dynamics community often focuses on specific applications; we suggest that system dynamics might increase its contribution to more generalized findings. System archetypes (Wolstenholme, 2003) help to generalize a model structure to multiple applications; it might also be possible to generalize findings from system dynamics into scientific learning more generally. Given the emphasis on causality in system dynamics (defining characteristic 1), the field might help to bridge theory and data, and to develop, compare, and combine theories in order to test their applicability to real-world systems. This would require extension of our results outside a specific application or the system dynamics field, perhaps putting more emphasis on incorporating our methods into the abductive reasoning process (Schwaninger and Hamann, 2005; Barton and Haslett, 2006) meant to develop generalized theory about the world itself.

Some effort has been put into defining the position of system dynamics within the larger scientific environment. Pruyt (2006), for example, examined system dynamics' relationship to the major theoretical research paradigms of positivism, constructivism, postpositivism, pragmatism, and criticalism, and found that different system dynamics practices fall within different paradigms. We suggest that research into system dynamics' place in the research community continue to be explored, but with a broader view that focuses less on how we think of system dynamics and more on how we can contribute. For example, how can system dynamics contribute to new discoveries and theories, how should it be integrated with other methods, and what components of its defining characteristics should those outside of our field begin to incorporate?

Historically, system dynamics has been a relatively slow-moving field, which has limited its impact. Improving our ability to quickly tackle new and urgent problems, either through generalizable models or through new methods of model development, could increase our impact on scientific understanding and decision making. Faster turnaround would likely necessitate new methods, perhaps incorporating advances in causality and data science mentioned above.

Research questions on how system dynamics can contribute to scientific advancement include:
- *How should system dynamics be integrated into broader scientific research?*
- *Which defining characteristics of system dynamics can help, or limit, generalized use of findings?*
- *Can faster turnaround increase the impact of system dynamics on urgent topics, and what new methods are needed to facilitate this?*

## Conclusions

In this article, we introduced the defining characteristics of system dynamics modeling, discussed those characteristics and their interactions, and presented a selection of aligned research opportunities. The defining characteristics are meant to give us a concrete shared understanding of what system dynamics modeling is, which models should be considered system dynamics models, and how the field can evolve while holding on to its core philosophies and practices. Previous definitions of the field have struggled to be clear, concise, and responsive to innovation.

The defining characteristics of system dynamics can help in education and communication about system dynamics to non-experts. System dynamics education opportunities are relatively limited, with only a few dedicated university programs, and much of the field's knowledge is translated through mentorship or informally held by its most consistent practitioners. A clear set of defining characteristics might help us to organize and clarify communication about the field. These characteristics can be used both at the high level, such as for brief topic introductions on social media platforms, and also more deeply in combination with other efforts to formalize the body of knowledge (Schaffernicht and Groesser, 2016; Arnold and Wade, 2017). They can also help to improve our communication with other fields, clarifying what we consider to be system dynamics and how that might dovetail with other paradigms. There may even be opportunities to explore how individual defining characteristics contribute to interpretation and understanding of system dynamics.

Defining characteristics might also help to create a more coherent and consistent identity for the system dynamics field. The current lack of clarity and resulting identity crisis (Homer 2013) is rooted in ambiguous philosophical and theoretical directions. In particular, there has been a disagreement about how system dynamics should be defined (Clancy, 2023). Größler (2013) noted that establishing a coherent identity for the field holds some danger due to its potential for limiting diversity; the defining characteristics presented here attempt to create identity while promoting diversity and innovation.

The list of opportunities presented here is by no means exhaustive. There are many more potential extensions of system dynamics modeling open for pursuit; for example, we have not discussed opportunities for improving spatial modeling capabilities or explored systems thinking. We encourage expansion of this list of opportunities by the broader community.

## Acknowledgements

Sandia National Laboratories is a multimission laboratory managed and operated by National Technology & Engineering Solutions of Sandia, LLC, a wholly owned subsidiary of Honeywell International Inc., for the U.S. Department of Energy's National Nuclear Security Administration under contract DE-NA0003525. This paper describes objective technical results and analysis. Any subjective views or opinions that might be expressed in the paper do not necessarily represent the views of the U.S. Department of Energy or the United States Government.